\documentclass[11pt]{article}
 
\usepackage{amsmath}
 \usepackage{tabularx}
 \usepackage{array}
 \usepackage{booktabs}
 \usepackage{graphicx}
 \usepackage{longtable}
 \def\CP{{$\cal CP~$}}
 \usepackage{amsfonts}
 \usepackage{float}
\usepackage{hyperref}
\usepackage{bm, geometry, cite, amssymb}
\begin{document}
 
 \title{\bf Correlation-Induced Interferometric Geometric Phase Difference in Entangled Neutral Meson Systems}

 \author{
 	{\bf Swarup Sangiri}\thanks{swarup.phys@gmail.com}\\[4pt]
 \small Physics Department, Indian Institute of Technology Kharagpur\\
	 \small Kharagpur 721302, India\\
}

\date{}
\maketitle
\begin{abstract}
We investigate the phase structure associated with correlated evolution in entangled neutral meson systems. Working within a rephasing-invariant interferometric framework, we construct the time-dependent interferometric geometric phase associated with the antisymmetric entangled neutral meson state undergoing nonunitary evolution. Exploiting the natural factorization of the overlap amplitude into global propagation and interference contributions, we derive an explicit expression for the phase in terms of the mass and decay-width differences that govern neutral meson mixing. To place the entangled phase in context, we derive the corresponding interferometric geometric phases accumulated by independently evolving neutral mesons within the same framework. This comparison naturally leads to the introduction of a correlation-induced interferometric geometric phase difference, $\Delta\gamma=\gamma_{\rm g}^{\rm{ent}}-\gamma_{\rm g}^{(1)}-\gamma_{\rm g}^{(2)}$, defined as the deviation of the entangled interferometric geometric phase from the sum of the associated single-meson phases. We show that this quantity characterizes the nonadditive phase structure generated by correlated meson evolution and originates from interference between propagation pathways that have no analogue in isolated meson dynamics. The resulting phases depend solely on the eigenvalue differences governing neutral meson oscillations and decay. The system-dependent behavior of the interferometric geometric phases and the correlation-induced phase difference is analyzed across different neutral meson families, illustrating how the underlying mixing dynamics shape the resulting phase correlations. Our results provide an interferometric characterization of entangled neutral meson evolution and highlight the role of entanglement in generating nonadditive phase structures associated with correlated meson dynamics.
\end{abstract}

\newpage

\section{Introduction}
Einstein, Podolsky, and Rosen (EPR) first recognized a ``spooky" feature of quantum mechanics \cite{Einstein1935}, highlighting the existence of strong correlations between spatially separated systems, later formalized through Bell inequalities \cite{Bell1964}. Erwin Schr\"{o}dinger, in his response to the EPR paper, first clearly identified and introduced the notion of quantum entanglement \cite{Schrodinger1935a,Schrodinger1935b,Schrodinger1935c}.

Entangled states arise naturally in particle physics. The first neutral pseudoscalar meson-antimeson system to exhibit entanglement was the $K^0-\overline{K^0}$ system \cite{GellMannPais1955, LeeOehmeYang1957}. Later mixing effects were also seen in $B_d^0-\overline{B_d^0}$ \cite{ARGUS1987}, $B_s^0-\overline{B_s^0}$ and $D^0-\overline{D^0}$ \cite{PDG2024}. These pseudoscalar systems provide a suitable platform for studying both phenomenological and experimental effects of quantum entanglement \cite{BertlmannHiesmayr2001, Horodecki2009}. Entangled neutral meson pairs are experimentally realized in $\phi$ and $\Upsilon(4S)$ decays as demonstrated by KLOE, BABAR and Belle collaborations \cite{KLOE2006, BABAR2004, Belle2001}.

At the microscopic level, discrete symmetries play a central role in understanding particle interactions. In particular, the combined charge-conjugation and parity symmetry, denoted by \CP, has long provided important insights into the structure of flavor dynamics. The discovery of \CP violation in neutral kaon decays \cite{Christenson1964} established neutral meson systems as a unique laboratory for studying mixing, oscillations, and symmetry breaking phenomena \cite{KobayashiMaskawa1973,BigiSanda2009,Sarkar2008}. The effective Hamiltonian governing neutral meson evolution contains complex mixing parameters that encode these dynamical features and determine the oscillation and decay properties of the system. From an interferometric perspective, the evolution generated by such mixing dynamics naturally motivates the investigation of phase structures associated with the relative propagation of neutral meson states and the interference effects induced by mixing and decay.

Geometric phases provide a powerful language for characterizing such systems. Originally introduced by Berry in the context of cyclic adiabatic evolution under a time-dependent Hermitian Hamiltonian \cite{Berry1984}, the concept has been extended to nonadiabatic, noncyclic and nonunitary evolution \cite{AharonovAnandan1987, SamuelBhandari1988, GarrisonWright1988}.  The development of geometric phase concepts has provided insight into fundamental structural properties of quantum mechanics \cite{MukundaSimon1993, ShapereWilczek1989}. Since entangled neutral meson systems evolve under an effective non-Hermitian Hamiltonian, the formulation of geometric phases for such systems require a generalized framework.
Recent studies have explored various aspects of quantum entanglement, geometric phases, and \CP-violating phenomena in meson systems and related quantum setups \cite{Xing2025, Dorigo2025,  Gabrielli2024, Chen2024, Nebot2017}.

Geometric phases have previously been investigated in single neutral meson systems, where the phase structure is associated with the relative evolution of the heavy and light mass-eigenstate components of an individual meson. In particular, geometric phases arising in neutral kaon mixing and their relation to indirect \CP-violating dynamics have been explored in earlier studies \cite{SangiriSarkar2023}. In the present work, we extend these ideas to entangled neutral meson pairs by constructing a rephasing-invariant interferometric framework for correlated two-meson evolution. Exploiting the natural factorization of the overlap amplitude into global propagation and interference contributions, we derive the time-dependent interferometric geometric phase associated with the entangled state and compare it with the corresponding phases accumulated by independently evolving mesons.

This comparison naturally leads to the introduction of a correlation-induced geometric phase difference,
\[
\Delta\gamma
=
\gamma_{\rm g}^{\rm{ent}}
-
\gamma_{\rm g}^{(1)}
-
\gamma_{\rm g}^{(2)},
\]
defined as the deviation of the entangled interferometric geometric phase from the sum of the corresponding single-meson phases. We show that this quantity characterizes the nonadditive phase structure generated by the correlated evolution of the entangled state and provides a direct measure of phase nonfactorization in the two-meson system. The resulting interferometric phases depend solely on the eigenvalue differences $\Delta m$ and $\Delta\Gamma$ that govern neutral meson mixing, thereby maintaining a direct connection with the standard oscillation formalism. In this way, the present work provides an interferometric characterization of correlated meson evolution and highlights aspects of the entangled phase structure that cannot be reconstructed from independent single-particle contributions alone.

\section{Generalized Neutral Meson Mixing: Formalism}
We briefly review the quantum mechanical formalism for neutral meson mixing and how \CP violation enters in the mixing dynamics. Consider a general particle-antiparticle ($P^0$-$\overline{P^0}$) mixing where the particles and antiparticles are distinguished by flavor quantum numbers such as strangeness, charm or beauty. Here $P^0$ can represent $K^0$, $D^0$ or $B^0$, and is distinguished from its antiparticle $\overline{P^0}$ by flavor quantum numbers that are conserved in strong and QED interactions but violated by weak interactions. Consequently, their flavor eigenstates are not stationary under weak interactions, thereby inducing mixing \cite{LeeOehmeYang1957}.

Within the Weisskopf--Wigner approximation, the time evolution of the neutral meson system is governed by the effective non-Hermitian Hamiltonian \cite{Waldi2001,BigiSanda2009}
\begin{align}
\mathcal{H}
=
\mathbf{M}
-
\frac{i}{2}\mathbf{\Gamma}
=
\begin{pmatrix}
M_{11}-\dfrac{i}{2}\Gamma_{11}
&
M_{12}-\dfrac{i}{2}\Gamma_{12}
\\[1ex]
M_{12}^{*}-\dfrac{i}{2}\Gamma_{12}^{*}
&
M_{22}-\dfrac{i}{2}\Gamma_{22}
\end{pmatrix},
\end{align}
where $\mathbf{M}$ and $\mathbf{\Gamma}$ are Hermitian matrices.

Assuming $\mathcal{CPT}$ invariance,
\begin{align}
M_{11}=M_{22}\equiv M,
\qquad
\Gamma_{11}=\Gamma_{22}\equiv\Gamma.
\end{align}

The mass eigenstates satisfy the eigenvalue equation
\begin{align}
\mathcal{H}|P_{H,L}\rangle
=
\lambda_{H,L}|P_{H,L}\rangle,
\end{align}
with complex eigenvalues
\begin{align}
\lambda_{H,L}
=
m_{H,L}
-
\frac{i}{2}\Gamma_{H,L},
\end{align}
where the subscripts $H$ and $L$ denote the heavy and light states, respectively.

The mass eigenstates are related to the flavor eigenstates through
\begin{align}
|P_{H,L}\rangle
=
p|P^0\rangle
\pm
q|\overline{P^0}\rangle,
\label{PHL}
\end{align}
with the normalization condition
\begin{align}
|p|^2+|q|^2=1,
\end{align}
and
\begin{align}
\frac{q}{p}
=
\sqrt{
\frac{
M_{12}^{*}-\dfrac{i}{2}\Gamma_{12}^{*}
}{
M_{12}-\dfrac{i}{2}\Gamma_{12}
}
}.
\end{align}

Throughout this work, we adopt the conventional phenomenological treatment in which the heavy and light mass eigenstates are assumed to form an orthonormal basis under the standard inner product,
\begin{align}
\langle P_i|P_j\rangle
=
\delta_{ij},
\qquad
i,j\in\{H,L\}.
\label{eq:orthonormality}
\end{align}

Since $P_{H,L}$ are eigenstates of $\cal H$, we can express their time evolution in the standard exponential form
\begin{align}
|P_{H,L}(t)\rangle
=
e^{-i\lambda_{H,L}t}
|P_{H,L}(0)\rangle.
\label{eq:time_evolution_P_H_L}
\end{align}

Inverting Eq.~(\ref{PHL}), the flavor eigenstates evolve according to
\begin{subequations}
\label{eq:time_evolution_P}
\begin{align}
|P^0(t)\rangle
&=
g_+(t)\,|P^0\rangle
+
\frac{q}{p}\,g_-(t)\,|\overline{P^0}\rangle,
\\
|\overline{P^0}(t)\rangle
&=
\frac{p}{q}\,g_-(t)\,|P^0\rangle
+
g_+(t)\,|\overline{P^0}\rangle,
\end{align}
\end{subequations}
where
\begin{align}
g_{\pm}(t)
=
\frac{1}{2}
\left(
e^{-i\lambda_H t}
\pm
e^{-i\lambda_L t}
\right).
\end{align}

The standard time-dependent function $g_+(t)$
describes the survival amplitude of the initial flavor state, while $g_-(t)$ describes flavor oscillations induced by mixing. 
If  \CP is conserved in mixing,
\begin{align}
\big|\frac{q}{p}\big|=1. 
\end{align}
When $|{q\over p}|\ne 1$, the heavy and light states are no longer \CP eigenstates, and the mixing dynamics distinguish between matter and antimatter at the propagation level. This phenomenon is known as indirect \CP violation, or \CP violation in mixing.

For later convenience, we define the mass and decay-width differences as
\begin{align}
\Delta m
&=
m_H-m_L,
\\
\Delta\Gamma
&=
\Gamma_H-\Gamma_L,
\end{align}
together with the average decay width
\begin{align}
\Gamma
=
\frac{\Gamma_H+\Gamma_L}{2}.
\end{align}

Since the effective Hamiltonian is non-Hermitian, the corresponding time evolution is nonunitary. In the following sections, we exploit the structure of the mass-eigenstate evolution to construct rephasing-invariant interferometric geometric phases for both entangled and independently evolving neutral mesons.

\section{Entangled Neutral Meson States}
Entangled neutral meson-antimeson pairs are produced in high-energy experiments through the decay of vector resonances formed in electron-positron collisions \cite{BigiSanda2009, KLOE2006, BABAR2004}. Examples include the decays $\phi(1020)\rightarrow K^0\overline{K^0}$ and $\Upsilon(4S)\rightarrow B^0\overline{B^0}$. Since the parent resonances carry total spin $J=1$ and charge conjugation eigenvalue $\mathcal{C}=-1$, the two-meson system is subject to the following constraints \cite{BertlmannHiesmayr2001}:

(a) The spatial part of the two-particle wavefunction is antisymmetric under particle exchange. Since the decay products are pseudoscalar mesons with zero intrinsic spin, conservation of angular momentum requires the two-meson system to be produced with relative orbital angular momentum $L=1$.

(b) The neutral meson pair inherits the charge-conjugation quantum number of the parent vector resonance. 

Thus, the initial state of the neutral meson pair at the time of creation is given by \cite{Lipkin1989, BertlmannGrimus1997}
\begin{align}
|\Psi(0)\rangle = \frac{1}{\sqrt{2}} \left( |P^0\rangle_1|\overline P^0\rangle_2 - |\overline P^0\rangle_1|P^0\rangle_2 \right),
\label{Psi_0_ent}
\end{align}
where $P^0$ represents a pseudoscalar meson like $K^0$, $D^0$ or $B^0$.  This state represents a maximally entangled Bell-type configuration in the flavor basis at production and cannot be factorized into independent single-meson states \cite{Bertlmann2001, Go2004}. As a direct consequence of the antisymmetric structure, the observation of a definite flavor for one meson immediately determines the flavor of its partner. 

The time evolution of the entangled meson pair can be written as 
\begin{align}
|\Psi(t_1,t_2)\rangle = \frac{1}{\sqrt{2}} \left( |P^0(t_1)\rangle_1|\overline P^0(t_2)\rangle_2 - |\overline P^0(t_1)\rangle_1|P^0(t_2)\rangle_2 \right),
\end{align}
where the subscripts label the spatially separated mesons. Under the simultaneous exchange of the meson labels and their corresponding evolution times, $(1,t_1)\leftrightarrow(2,t_2)$, the state remains antisymmetric, $|\Psi(t_1,t_2)\rangle = -|\Psi(t_2,t_1)\rangle$.

Observable consequences of entanglement can be seen in joint decay processes, where both mesons decay into specified final states at different times. The corresponding joint decay intensity 
$I(f_1,t_1; f_2,t_2)\propto |\langle f_1, f_2 |\Psi(t_1,t_2)\rangle |^2$ depends not only on the individual decay amplitudes, but also on the interference terms generated by the superposition of flavor-transition pathways \cite{BABAR2004, Belle2001}. In subsequent sections, we show that these entangled neutral meson pairs constitute a natural interferometric system for investigating the phase structure associated with correlated neutral meson evolution. In particular, we construct rephasing-invariant interferometric geometric phases and examine how their nonadditive behavior is governed by the underlying mixing dynamics.

\section{Time-Dependent interferometric Geometric Phase of the Entangled State}

Geometric phases in neutral meson systems have been investigated as potential probes of mixing-induced interference and \CP violation. In particular, it has been shown that, within the Berry phase framework, \CP-violating effects in the neutral kaon system acquire a geometric interpretation \cite{Sangiri2024}. In this section, we formulate an interferometric geometric phase associated with the correlated evolution of an entangled neutral meson system undergoing nonunitary evolution.

To evaluate the phase accumulated during the evolution of the entangled system, we consider the initial state at $t=0$:
\begin{align}
|\Psi(0)\rangle
=
\frac{1}{\sqrt{2}}
\left(
|P^0\rangle_1 |\overline P^0\rangle_2
-
|\overline P^0\rangle_1 |P^0\rangle_2
\right),
\end{align}
where $1,2$ label the two spatially separated mesons. Using the inversion of Eq.~(\ref{PHL}), we obtain
\begin{align}
|\Psi(0)\rangle
=
\frac{1}{\sqrt{2}}
\frac{1}{2pq}
\left(
|P_L\rangle_1 |P_H\rangle_2
-
|P_H\rangle_1 |P_L\rangle_2
\right).
\end{align}

Using Eq.~(\ref{eq:time_evolution_P_H_L}), the time-evolved two-particle state at $(t_1,t_2)$ is
\begin{align}
|\Psi(t_1,t_2)\rangle
=
\frac{1}{\sqrt{2}}
\frac{1}{2pq}
\left(
e^{-i\lambda_L t_1-i\lambda_H t_2}
|P_L\rangle_1 |P_H\rangle_2
-
e^{-i\lambda_H t_1-i\lambda_L t_2}
|P_H\rangle_1 |P_L\rangle_2
\right).
\end{align}

The overlap amplitude associated with the correlated evolution is defined as
\begin{align}
\mathcal{A}(t_1,t_2)
=
\langle\Psi(0)|\Psi(t_1,t_2)\rangle .
\end{align}

In the present work, we adopt the conventional phenomenological treatment of neutral meson mixing in which the heavy and light mass eigenstates are assumed to form an orthonormal basis under the standard inner product. Within this framework, we obtain
\begin{align}
\mathcal{A}(t_1,t_2)
=
\frac{1}{8|p|^2|q|^2}
\left(
e^{-i\lambda_H t_2-i\lambda_L t_1}
+
e^{-i\lambda_L t_2-i\lambda_H t_1}
\right).
\end{align}

This may be rewritten as
\begin{align}
\mathcal{A}(t_1,t_2)
=
\frac{1}{4|p|^2|q|^2}
e^{-i\frac{\lambda_H+\lambda_L}{2}(t_1+t_2)}
\cos\!\left[
\frac{\lambda_H-\lambda_L}{2}(t_2-t_1)
\right].
\end{align}

This form is particularly useful because it factorizes into a global propagation term depending on $(t_1+t_2)$ and an interference term depending on the time difference $(t_2-t_1)$.

The total phase accumulated by the entangled system is defined through the overlap amplitude as
\begin{align}
\gamma^{\rm ent}_{\rm{tot}}(t_1,t_2)
=
\arg \mathcal{A}(t_1,t_2).
\end{align}

Using $\lambda_{H,L}=m_{H,L}-\frac{i}{2}\Gamma_{H,L}$, we obtain
\begin{align}
\gamma^{\rm ent}_{\rm tot}
=
-\frac12(m_H+m_L)(t_1+t_2)
+
\arg\!\left[
\cos\!\left(
\frac{\Delta m}{2}(t_2-t_1)
-
i\frac{\Delta\Gamma}{4}(t_2-t_1)
\right)
\right].
\end{align}

The overlap amplitude naturally factorizes into a global propagation term and an interference term. The exponential factor $e^{-i\frac{\lambda_H+\lambda_L}{2}(t_1+t_2)}$ describes the overall propagation of the bipartite state under the effective Hamiltonian, whereas the cosine factor encodes the interference between the two indistinguishable propagation pathways $e^{-i\lambda_Lt_1-i\lambda_Ht_2}$ and $e^{-i\lambda_Ht_1-i\lambda_Lt_2}$. Motivated by this natural decomposition, we define an interferometric geometric phase as the residual phase obtained after removing the contribution associated with the global propagation factor. Within the present framework, the phase of the exponential term, 
\begin{align}
    \gamma_{\rm prop}^{\rm ent}=-\frac{1}{2}(m_H+m_L)(t_1+t_2),
\end{align}
is identified as the propagation phase.

Following the standard decomposition $\gamma_{\rm g}^{\rm ent}\equiv\gamma_{\rm tot}^{\rm ent}-\gamma_{\rm prop}^{\rm ent}$, the interferometric geometric phase of the entangled neutral meson system is obtained by subtracting the global propagation contribution from the total phase:
\begin{align}
\gamma_{\rm g}^{\rm ent}
=
\gamma^{\rm ent}_{\rm tot}
-
\gamma^{\rm ent}_{\rm prop}
=
\arg\!\left[
\cos\!\left(
\frac{\Delta m}{2}(t_2-t_1)
-
i\frac{\Delta\Gamma}{4}(t_2-t_1)
\right)
\right].
\label{eq:gp_ent}
\end{align}

The resulting phase is therefore associated with the interference structure of the correlated two-meson evolution rather than with the overall propagation of the state. Since it is obtained after removing the global propagation contribution, it provides an interferometric characterization of the correlated dynamics of the entangled meson pair.

Using $\cos(a-ib)=\cos a\,\cosh b+i\,\sin a\,\sinh b$, with $a=\frac{\Delta m}{2}(t_2-t_1)$ and $b=\frac{\Delta\Gamma}{4}(t_2-t_1)$, the interferometric geometric phase may be written explicitly as
\begin{align}
\gamma_{\rm g}^{\rm ent}
=
\tan^{-1}
\!\left[
\frac{
\sin\!\left(\frac{\Delta m}{2}\Delta t\right)
\sinh\!\left(\frac{\Delta\Gamma}{4}\Delta t\right)
}{
\cos\!\left(\frac{\Delta m}{2}\Delta t\right)
\cosh\!\left(\frac{\Delta\Gamma}{4}\Delta t\right)
}
\right],
\end{align}
where $\Delta t=t_2-t_1$. Equivalently,
\begin{align}
\gamma_{\rm g}^{\rm ent}
=
\tan^{-1}
\!\left[
\tan\!\left(
\frac{\Delta m}{2}\Delta t
\right)
\tanh\!\left(
\frac{\Delta\Gamma}{4}\Delta t
\right)
\right].
\label{eq:gp_ent_explicit}
\end{align}

The principal branch of the inverse tangent is assumed throughout, with the phase understood modulo $2\pi$. Branch discontinuities may therefore occur when the denominator of Eq.~(\ref{eq:gp_ent_explicit}) vanishes. This representation makes the interplay between the oscillation scale $\Delta m$ and the decay-width difference $\Delta\Gamma$ explicit and will be particularly useful when comparing the entangled geometric phase with the geometric phases accumulated by independently evolving neutral mesons.

The interferometric geometric phase derived above is invariant under independent rephasings of the flavor states:
\begin{align}
|P^0\rangle \rightarrow e^{i\alpha}|P^0\rangle,
\qquad
|\overline P^0\rangle \rightarrow e^{i\beta}|\overline P^0\rangle.
\end{align}

Under this transformation the mixing parameters transform as
\begin{align}
p\rightarrow e^{i\alpha}p,
\qquad
q\rightarrow e^{i\beta}q,
\end{align}
while the eigenvalues of the effective Hamiltonian, $\lambda_{H,L}$, remain unchanged. The entangled state transforms by an overall phase,
\begin{align}
|\Psi(0)\rangle
\rightarrow
e^{i(\alpha+\beta)}
|\Psi(0)\rangle.
\end{align}

Since Eqs.~(\ref{eq:gp_ent}) and (\ref{eq:gp_ent_explicit}) depend only on the rephasing-invariant combinations $\lambda_H+\lambda_L$ and $\lambda_H-\lambda_L$, the resulting interferometric geometric phase $\gamma_{\rm g}^{\rm ent}$ is manifestly invariant under independent rephasings of the flavor states.

The interferometric geometric phase obtained in Eqs.~(\ref{eq:gp_ent}) and (\ref{eq:gp_ent_explicit}) is governed solely by the eigenvalue differences $\Delta m$ and $\Delta \Gamma$ of the effective Hamiltonian. Since these quantities determine the oscillation and decay properties of the neutral meson system, the phase remains directly connected to the underlying mixing dynamics. Within the present framework, no explicit dependence on the mixing parameter $q/p$ appears in the final expression. Consequently, the phase probes the interference associated with the relative propagation of the heavy and light mass eigenstates rather than indirect \CP-violating effects encoded in the mixing parameter. Unlike the global propagation contribution, which depends on the average eigenvalue $\frac{\lambda_H+\lambda_L}{2}$, the interferometric geometric phase is governed entirely by the relative evolution of the two mass eigenstates.

Having established the interferometric geometric phase associated with the entangled two-meson state, it is instructive to compare it with the geometric phases accumulated by independently evolving neutral mesons. This comparison will allow us to investigate whether the phase structure of the correlated bipartite system can be decomposed into a sum of single-particle contributions and to identify possible nonadditive features arising from entanglement.

\section{Interferometric Geometric Phases of Independently Evolving Neutral Mesons}

To assess the role of entanglement in the phase structure of the bipartite system, it is useful to compare the interferometric geometric phase obtained in the previous section with the corresponding phases accumulated by independently evolving neutral mesons. We therefore consider a single neutral meson prepared initially in a flavor eigenstate and evolving under the same effective Hamiltonian.

Starting from the flavor state $|P^0\rangle$, we use the inverse transformation of Eq.~(\ref{PHL}) to write
\begin{align}
|P^0\rangle
=
\frac{1}{2p}
\left(
|P_H\rangle
+
|P_L\rangle
\right).
\end{align}

Using Eq.~(\ref{eq:time_evolution_P_H_L}), the time-evolved state is
\begin{align}
|P^0(t)\rangle
=
\frac{1}{2p}
\left(
e^{-i\lambda_H t}|P_H\rangle
+
e^{-i\lambda_L t}|P_L\rangle
\right).
\end{align}

The corresponding overlap amplitude is $\mathcal{A}_1(t)=\langle P^0|P^0(t)\rangle$. Within the effective description adopted in the previous section, where the heavy and light mass eigenstates are assumed to form an orthonormal basis under the standard inner product, 
we obtain
\begin{align}
\mathcal{A}_1(t)
=
\frac{1}{4|p|^2}
\left(
e^{-i\lambda_H t}
+
e^{-i\lambda_L t}
\right).
\end{align}

Factoring out the average propagation term gives
\begin{align}
\mathcal{A}_1(t)
=
\frac{1}{2|p|^2}
e^{-i\frac{\lambda_H+\lambda_L}{2}t}
\cos\!\left[
\frac{\lambda_H-\lambda_L}{2}t
\right].
\end{align}

The total phase accumulated by the single meson is therefore $\gamma^{(1)}_{\rm tot}=\arg \mathcal{A}_1(t)$, or explicitly,
\begin{align}
\gamma^{(1)}_{\rm tot}
=
-\frac12(m_H+m_L)t
+
\arg\!\left[
\cos\!\left(
\frac{\Delta m}{2}t
-
i\frac{\Delta\Gamma}{4}t
\right)
\right].
\end{align}

Within the present interferometric framework, the propagation phase is identified with the phase generated by the global propagation factor,
\begin{align}
\gamma^{(1)}_{\rm prop}
=
-\frac12(m_H+m_L)t.
\end{align}

Motivated by the natural factorization of the overlap amplitude into a global propagation term and an interference term, we define the interferometric geometric phase of the single-meson system as the residual phase obtained after removing the contribution associated with the global propagation factor:
\begin{align}
\gamma_{\rm g}^{(1)}
=
\gamma^{(1)}_{\rm tot}
-
\gamma^{(1)}_{\rm prop}
=
\arg\!\left[
\cos\!\left(
\frac{\Delta m}{2}t
-
i\frac{\Delta\Gamma}{4}t
\right)
\right].
\label{eq:gp_single}
\end{align}

Using $\cos(a-ib)=\cos a\,\cosh b+i\,\sin a\,\sinh b$, the interferometric geometric phase may be written explicitly as
\begin{align}
\gamma_{\rm g}^{(1)}
=
\tan^{-1}
\!\left[
\tan\!\left(
\frac{\Delta m}{2}t
\right)
\tanh\!\left(
\frac{\Delta\Gamma}{4}t
\right)
\right].
\label{eq:gp_single_explicit}
\end{align}

As in the entangled-state case, the principal branch of the inverse tangent is assumed, with the phase understood modulo $2\pi$. An identical analysis applies to the second meson evolving independently for a time $t_2$. The corresponding geometric phase is therefore
\begin{align}
\gamma_{\rm g}^{(2)}
=
\tan^{-1}
\!\left[
\tan\!\left(
\frac{\Delta m}{2}t_2
\right)
\tanh\!\left(
\frac{\Delta\Gamma}{4}t_2
\right)
\right].
\label{eq:gp_single2}
\end{align}

Comparing Eqs.~(\ref{eq:gp_ent_explicit}), (\ref{eq:gp_single_explicit}) and (\ref{eq:gp_single2}), one observes that the functional dependence of the interferometric geometric phase is identical in all three cases. The essential difference lies in the relevant time variable: the entangled geometric phase depends on the relative evolution time $\Delta t=t_2-t_1$, whereas the independently evolving mesons accumulate phases according to their individual propagation times. This distinction motivates a direct comparison between the correlated bipartite phase and the sum of the corresponding single-particle geometric phases. Such a comparison will allow us to examine whether the phase structure of the entangled system can be reconstructed from independent single-particle contributions or whether it exhibits genuinely nonadditive features associated with correlated evolution.

\section{Correlation-Induced Interferometric Geometric Phase Difference}

Having derived the interferometric geometric phase associated with the entangled neutral meson system and the corresponding phases accumulated by independently evolving mesons, we now investigate the extent to which the phase structure of the correlated bipartite system can be decomposed into a sum of single-particle contributions.

A natural quantity for this purpose is the difference
\begin{align}
\Delta\gamma
=
\gamma_{\rm g}^{\rm ent}
-
\gamma_{\rm g}^{(1)}
-
\gamma_{\rm g}^{(2)},
\label{eq:deltagamma_def}
\end{align}
which measures the deviation of the interferometric geometric phase of the entangled system from the sum of the corresponding phases accumulated by two independently evolving mesons.

Using Eqs.~(\ref{eq:gp_ent_explicit}), (\ref{eq:gp_single_explicit}) and (\ref{eq:gp_single2}), we obtain
\begin{align}
\Delta\gamma
&=
\tan^{-1}
\!\left[
\tan\!\left(
\frac{\Delta m}{2}(t_2-t_1)
\right)
\tanh\!\left(
\frac{\Delta\Gamma}{4}(t_2-t_1)
\right)
\right]
\nonumber\\[1ex]
&\quad
-
\tan^{-1}
\!\left[
\tan\!\left(
\frac{\Delta m}{2}t_1
\right)
\tanh\!\left(
\frac{\Delta\Gamma}{4}t_1
\right)
\right]
\nonumber\\[1ex]
&\quad
-
\tan^{-1}
\!\left[
\tan\!\left(
\frac{\Delta m}{2}t_2
\right)
\tanh\!\left(
\frac{\Delta\Gamma}{4}t_2
\right)
\right].
\label{eq:deltagamma_exact}
\end{align}

Throughout this section the principal branch of the inverse tangent is assumed, with geometric phases understood modulo $2\pi$. Equation~(\ref{eq:deltagamma_exact}) provides an exact measure of the nonadditivity of interferometric geometric phase accumulation in the correlated meson system. By construction, $\Delta\gamma$ vanishes whenever the interferometric geometric phase of the bipartite state can be represented as a simple sum of single-particle contributions. 

To further clarify the origin of the nonadditive contribution, it is useful to consider a general separable bipartite state,
\begin{align}
|\Psi(t_1,t_2)\rangle
=
|\psi_1(t_1)\rangle
\otimes
|\psi_2(t_2)\rangle,
\end{align}
with corresponding initial state
\begin{align}
|\Psi(0,0)\rangle
=
|\psi_1(0)\rangle
\otimes
|\psi_2(0)\rangle .
\end{align}

The overlap amplitude associated with the evolution of this state is
\begin{align}
\mathcal A_{\rm sep}(t_1,t_2)
=
\langle \Psi(0,0)|\Psi(t_1,t_2)\rangle .
\end{align}
Using the tensor-product structure of the state, the overlap amplitude factorizes,
\begin{align}
\mathcal A_{\rm sep}(t_1,t_2)
&=
\langle\psi_1(0)|\psi_1(t_1)\rangle
\,
\langle\psi_2(0)|\psi_2(t_2)\rangle
\nonumber\\
&=
\mathcal A_1(t_1)\mathcal A_2(t_2).
\end{align}

The total phase therefore satisfies
\begin{align}
\gamma_{\rm tot}^{\rm sep}
&=
\arg\!\left[
\mathcal A_1(t_1)\mathcal A_2(t_2)
\right]
\nonumber\\
&=
\gamma_{\rm tot}^{(1)}
+
\gamma_{\rm tot}^{(2)}
\qquad
(\mathrm{mod}\;2\pi).
\end{align}

The propagation phase is likewise additive. Since the overlap amplitude factorizes, the corresponding global propagation factors also separate into independent contributions associated with the two subsystems. Consequently,
\begin{align}
\gamma_{\rm prop}^{\rm sep}
=
\gamma_{\rm prop}^{(1)}
+
\gamma_{\rm prop}^{(2)}.
\end{align}

Within the interferometric framework adopted in this work, the geometric phase is defined as the residual phase obtained after subtracting the contribution associated with the global propagation factor. The interferometric geometric phase of an arbitrary separable pure bipartite state is therefore additive,
\begin{align}
\gamma_{\rm g}^{\rm sep}
&=
\gamma_{\rm tot}^{\rm sep}
-
\gamma_{\rm prop}^{\rm sep}
\nonumber\\
&=
\gamma_{\rm g}^{(1)}
+
\gamma_{\rm g}^{(2)},
\end{align}
and hence
\begin{align}
\Delta\gamma_{\rm sep}
&=
\gamma_{\rm g}^{\rm sep}
-
\gamma_{\rm g}^{(1)}
-
\gamma_{\rm g}^{(2)}
\nonumber\\
&=
0.
\end{align}

The preceding derivation shows that interferometric geometric phase additivity holds for any factorized pure bipartite state. The nonzero value of $\Delta\gamma$ in Eq.~(\ref{eq:deltagamma_exact}) therefore originates from the nonfactorizable overlap amplitude of the entangled meson state, which prevents the total interferometric geometric phase from being decomposed into a sum of single-particle contributions. In particular, the overlap amplitude of the entangled system contains the two indistinguishable propagation pathways $e^{-i\lambda_L t_1-i\lambda_H t_2}$ and $e^{-i\lambda_H t_1-i\lambda_L t_2}$, whose interference has no counterpart in a separable evolution. A nonzero value of $\Delta\gamma$ therefore reflects the nonadditive phase structure associated with the correlated evolution of the entangled neutral meson pair.

To gain analytical insight, we consider the regime of small evolution times. Using
\begin{align}
\tan x \simeq x,
\qquad
\tanh y \simeq y,
\end{align}
one finds
\begin{align}
\gamma_{\rm g}^{\rm ent}
&\simeq
\frac{\Delta m\,\Delta\Gamma}{8}
(t_2-t_1)^2,
\\
\gamma_{\rm g}^{(1)}
&\simeq
\frac{\Delta m\,\Delta\Gamma}{8}
t_1^2,
\\
\gamma_{\rm g}^{(2)}
&\simeq
\frac{\Delta m\,\Delta\Gamma}{8}
t_2^2.
\end{align}

Substituting these expressions into Eq.~(\ref{eq:deltagamma_def}) yields
\begin{align}
\Delta\gamma
\simeq
-\frac{\Delta m\,\Delta\Gamma}{4}
t_1 t_2.
\label{eq:deltagamma_small}
\end{align}

Equation~(\ref{eq:deltagamma_small}) shows explicitly that the leading nonadditive contribution arises from the product of the two propagation times and therefore has no analogue in independently evolving single-meson systems. This result reveals an important structural feature of the correlated evolution. Unlike the individual interferometric geometric phases, which depend on a single propagation time, the leading contribution to $\Delta\gamma$ is proportional to the product $t_1 t_2$. Consequently, it cannot be attributed to either meson separately but instead depends simultaneously on the evolution histories of both constituents of the entangled pair.

Several limiting cases follow immediately from Eq.~(\ref{eq:deltagamma_small}). First, $\Delta\gamma \rightarrow 0$ for $t_1\rightarrow 0$ or $t_2\rightarrow 0$. In this limit the overlap amplitude depends effectively on the evolution of only one constituent, and the leading correlation-induced contribution vanishes. Similarly, $\Delta\gamma \rightarrow 0$ for $\Delta m \rightarrow 0$, showing that the effect vanishes when oscillations are absent. Furthermore, $\Delta\gamma \rightarrow 0$ for $\Delta\Gamma \rightarrow 0$, indicating that the geometric phase difference also disappears when the heavy and light states become indistinguishable through their decay widths.

These limits demonstrate that $\Delta\gamma$ is governed by the same eigenvalue differences that characterize neutral meson mixing, while simultaneously encoding information about the correlated nature of the bipartite evolution. The quantity therefore combines two distinct ingredients: the oscillation dynamics contained in $(\Delta m,\Delta\Gamma)$ and the nonfactorizable phase structure associated with the entangled state.

It is important to emphasize that $\Delta\gamma$ does not introduce a new mixing parameter and does not modify the conventional description of neutral meson oscillations. Rather, it provides an interferometric characterization of how the phase accumulated by the correlated two-meson system differs from that obtained by combining the phases of independently evolving mesons. In this sense, $\Delta\gamma$ may be viewed as a correlation-induced interferometric geometric phase difference associated with entangled neutral meson evolution.

\section{System Dependent Behavior of the Correlation-Induced interferometric Geometric Phase Difference}

In the previous section, we introduced the correlation-induced interferometric geometric phase difference
\begin{align}
\Delta\gamma
=
\gamma_{\rm g}^{\rm ent}
-
\gamma_{\rm g}^{(1)}
-
\gamma_{\rm g}^{(2)},
\end{align}
which quantifies the departure of the interferometric geometric phase accumulated by the entangled neutral meson system from the sum of the corresponding interferometric geometric phases accumulated by independently evolving mesons. We now investigate how this quantity behaves across different neutral meson families.

To facilitate comparison between systems with very different lifetimes and oscillation frequencies, it is convenient to introduce the dimensionless mixing parameters
\begin{align}
x=\frac{\Delta m}{\Gamma},
\qquad
y=\frac{\Delta\Gamma}{2\Gamma},
\end{align}
following the conventional normalization used in neutral meson phenomenology \cite{PDG2024}. Here
\begin{align}
\Delta m = m_H-m_L,
\qquad
\Delta\Gamma = \Gamma_H-\Gamma_L,
\qquad
\Gamma=\frac{\Gamma_H+\Gamma_L}{2}.
\end{align}
Our definition of
\(
y=(\Gamma_H-\Gamma_L)/(2\Gamma)
\)
differs by an overall sign from the Particle Data Group convention, however, this sign difference does not affect the qualitative discussion presented below.

Using the dimensionless variable $\Delta\tau=\Gamma(t_2-t_1)$, the interferometric geometric phase of Eq.~(\ref{eq:gp_ent_explicit}) becomes
\begin{align}
\gamma_{\rm g}^{\rm ent}
=
\tan^{-1}
\!\left[
\tan\!\left(
\frac{x}{2}\Delta\tau
\right)
\tanh\!\left(
\frac{y}{2}\Delta\tau
\right)
\right].
\end{align}

Similarly, the independently evolving mesons accumulate interferometric geometric phases
\begin{align}
\gamma_{\rm g}^{(i)}
=
\tan^{-1}
\!\left[
\tan\!\left(
\frac{x}{2}\tau_i
\right)
\tanh\!\left(
\frac{y}{2}\tau_i
\right)
\right],
\end{align}
where
\(
\tau_i=\Gamma t_i
\).
The correlation-induced geometric phase difference is then determined by Eq.~(\ref{eq:deltagamma_exact}).

The magnitude of $\Delta\gamma$ is controlled by the same mixing parameters $x$ and $y$ that govern the interferometric geometric phases themselves. Consequently, different neutral meson systems occupy distinct geometric regimes. In the following discussion we compare the systems at dimensionless evolution times of order unity, $\tau_i=\Gamma t_i \sim \mathcal O(1)$, corresponding to the physically relevant interference region in which oscillation and decay effects are both appreciable.

For the neutral kaon system,
\(
x_K\sim 1
\)
and
\(
|y_K|\sim 1
\)
\cite{PDG2024}.
In the physically relevant interference region
\(
\Delta\tau\sim\mathcal O(1)
\),
both the oscillatory and hyperbolic contributions remain unsuppressed. The entangled phase therefore acquires a sizable value, and the resulting $\Delta\gamma$ can be expected to exhibit the strongest deviation from additive single-particle behavior. The kaon system is expected to provide a particularly favorable environment for observing correlation-induced interferometric geometric effects.

For the $D^0$ system,
\(
x_D\sim10^{-3}
\)
and
\(
|y_D|\sim10^{-2}
\)
\cite{PDG2024}.
In this regime both quantities are small, and the geometric phases admit a systematic expansion. From Eq.~(\ref{eq:deltagamma_small}), one finds
\begin{align}
\Delta\gamma
\simeq
-\frac{x_Dy_D}{4}\,
\tau_1\tau_2,
\end{align}
to leading order. The resulting geometric correlation is therefore strongly suppressed.

For the $B_d^0$ system,
\(
x_d\sim\mathcal O(1)
\)
while
\(
|y_d|\ll1
\)
\cite{PDG2024}.
Since the leading contribution to the correlation-induced phase difference is proportional to the width-difference parameter, the geometric correlation becomes highly suppressed despite the sizeable oscillation frequency. In this limit the entangled and independent phase structures approach one another.

The situation is different for the $B_s^0$ system, where
\(
x_s\gg1
\)
and
\(
|y_s|\sim10^{-2}
\)
\cite{PDG2024}.
The large value of $x_s$ produces rapid oscillatory modulation, while the comparatively small value of $y_s$ controls the overall amplitude of the geometric correlation. Consequently, the overall magnitude of the geometric correlation is constrained by the comparatively small value of $y_s$, although the large oscillation frequency produces a rich oscillatory structure.

These observations indicate that a sizable width difference enhances the correlation-induced geometric phase difference, whereas the limit
\(
\Delta\Gamma\rightarrow0
\)
suppresses it almost entirely. The kaon system therefore appears to be the most promising neutral meson system for exhibiting a pronounced nonadditive geometric phase structure.

The qualitative geometric regimes for representative neutral meson systems are summarized in Table~\ref{tab:systemcomparison}.

\begin{table}[H]
\centering
\renewcommand{\arraystretch}{1.2}
\begin{tabular}{lccc}
\toprule
System & $|x|$ & $|y|$ & Geometric behavior\\
\midrule
$K^0$ & $\sim1$ & $\sim1$
& Expected strong nonadditive correlations\\

$D^0$ & $\sim10^{-3}$ & $\sim10^{-2}$
& Strongly suppressed\\

$B_d^0$ & $\sim1$ & $\sim0$
& Approximately additive\\

$B_s^0$ & $\sim10^1$ & $\sim10^{-2}$
& Oscillatory with suppressed amplitude\\
\bottomrule
\end{tabular}
\caption{Representative values of the mixing parameters and the expected behavior of the correlation-induced interferometric geometric phase difference $\Delta\gamma$.}
\label{tab:systemcomparison}
\end{table}

\section{Interferometric Phase Structure of Correlated Meson Evolution}

The results obtained in the preceding sections allow us to characterize the interferometric phase structure of correlated neutral meson evolution from a perspective that is distinct from the conventional description based solely on mixing amplitudes and decay probabilities.

For an independently evolving neutral meson, the interferometric geometric phase is associated with the relative evolution of the heavy and light mass eigenstate components of the flavor state. In this case, the phase is completely determined by the evolution of the individual meson and depends only on the eigenvalue differences $\Delta m$ and $\Delta\Gamma$ governing neutral meson mixing. The corresponding geometric phases $\gamma_{\rm g}^{(1)}$ and $\gamma_{\rm g}^{(2)}$ therefore characterize single-particle evolution.

In contrast, the entangled state evolves in the tensor-product Hilbert space of two flavor systems. The interferometric geometric phase $\gamma_{\rm g}^{\rm ent}$ is associated with the correlated evolution of the bipartite state and depends on the interference between the two indistinguishable propagation pathways $e^{-i\lambda_L t_1-i\lambda_H t_2}$ and $e^{-i\lambda_H t_1-i\lambda_L t_2}$, which have no analogue in the evolution of an isolated meson. Consequently, the interferometric geometric phase of the entangled system is not determined solely by the evolution of either subsystem but instead reflects the structure of the correlated two-particle state.

This distinction becomes particularly transparent through the quantity
\[
\Delta\gamma
=
\gamma_{\rm g}^{\rm ent}
-
\gamma_{\rm g}^{(1)}
-
\gamma_{\rm g}^{(2)},
\]
which measures the departure of the interferometric geometric phase of the entangled system from a simple sum of single-particle interferometric geometric phases. By construction, $\Delta\gamma$ vanishes whenever the geometric phase of the bipartite system can be reconstructed from independent subsystem contributions. A nonzero value therefore reflects the presence of phase correlations intrinsic to the entangled evolution.

The small-time behavior derived in Eq.~(\ref{eq:deltagamma_small}),
$\Delta\gamma=-\frac{\Delta m\,\Delta\Gamma}{4}t_1 t_2$, provides a particularly simple illustration of this feature. The leading contribution depends simultaneously on both propagation times and therefore cannot be associated with either meson individually. The appearance of the bilocal structure $t_1t_2$ reflects the correlated nature of the bipartite evolution and provides an interferometric manifestation of the nonfactorizable phase structure generated by the entangled state.

From this perspective, $\Delta\gamma$ may be viewed as an interferometric indicator of nonfactorization in the phase structure of the neutral meson system. Its existence originates from the correlated interference terms present in the entangled state and not from any modification of the underlying mixing formalism. The quantity therefore complements the conventional description of neutral meson evolution by providing a geometric characterization of the phase correlations generated by entanglement.

The role of $\Delta m$ and $\Delta\Gamma$ in Eqs.~(\ref{eq:gp_ent_explicit}) and (\ref{eq:deltagamma_exact}) further shows that the correlation-induced geometric phase difference is controlled by the same eigenvalue structure that governs flavor oscillations and lifetime differences. In this sense, the present interferometric framework remains firmly connected to the standard theory of neutral meson mixing while highlighting aspects of the correlated evolution that are not visible from the individual single-meson phases alone.

The various geometric quantities introduced in this work and their interpretations are summarized in Table~\ref{tab:geometricquantities}.

\begin{table}[H]
\centering
\renewcommand{\arraystretch}{1.2}
\begin{tabular}{p{3.2cm}p{9.5cm}}
\toprule
\textbf{Quantity} & \textbf{Interpretation} \\
\midrule

$\gamma_{\rm g}^{(1)}$
&
Interferometric geometric phase accumulated by a single independently evolving neutral meson.
\\

$\gamma_{\rm g}^{(2)}$
&
Interferometric geometric phase accumulated by the second independently evolving neutral meson.
\\

$\gamma_{\rm g}^{\rm ent}$
&
Interferometric geometric phase associated with the correlated evolution of the entangled two-meson state.
\\

$\Delta\gamma$
&
Correlation-induced interferometric geometric phase difference quantifying the departure from additive single-particle phase accumulation.
\\

Small-time limit:
$\Delta\gamma
\simeq
-\frac{\Delta m\Delta\Gamma}{4}t_1t_2$
&
Leading-order manifestation of interferometric phase nonfactorization; depends simultaneously on the evolution histories of both mesons.
\\

\bottomrule
\end{tabular}
\caption{Summary of the interferometric geometric quantities introduced in this work and their physical interpretation.}
\label{tab:geometricquantities}
\end{table}

\section{Summary}
In this work, we developed a rephasing-invariant framework for analyzing interferometric geometric phases in entangled neutral meson systems undergoing mixing and decay. Within the conventional phenomenological description of neutral meson mixing, we constructed the time-dependent interferometric geometric phase associated with the antisymmetric entangled neutral meson state by separating the global propagation contribution from the interference structure of the correlated evolution. The resulting phase is expressed entirely in terms of the eigenvalue differences governing neutral meson evolution.

To place the entangled interferometric geometric phase in context, we derived the corresponding phases accumulated by independently evolving neutral mesons within the same framework. This comparison allowed us to introduce a correlation-induced geometric phase difference, $\Delta\gamma= \gamma_{\rm g}^{\rm ent}-\gamma_{\rm g}^{(1)}-\gamma_{\rm g}^{(2)}$, which quantifies the deviation of the interferometric geometric phase of the entangled system from the sum of the associated single-particle geometric phases.

We showed that the interferometric geometric phase of the correlated two-meson state is generally nonadditive and that the quantity $\Delta\gamma$ provides a natural measure of this nonfactorization. The origin of this effect can be traced to interference terms that arise only in the entangled system and have no analogue in the evolution of an isolated meson. In the small-time limit, the leading contribution to $\Delta\gamma$ is proportional to $t_1t_2$, demonstrating that it depends simultaneously on the evolution histories of both mesons and therefore reflects the correlated phase structure of the entangled system.

The system-dependent behavior of the geometric phases was analyzed using the conventional dimensionless mixing parameters $x=\frac{\Delta m}{\Gamma}$, $y=\frac{\Delta\Gamma}{2\Gamma}$. We found that the magnitude of the correlation-induced interferometric geometric phase difference is controlled by the same eigenvalue differences that govern neutral meson oscillations and lifetime differences, with the kaon system providing the most favorable regime for sizable nonadditive phase correlations.

The framework developed here does not introduce new mixing parameters or modify the conventional phenomenology of neutral meson oscillations. Rather, it reorganizes the standard dynamical information into interferometric phase quantities that distinguish correlated and independent evolution. In this way, the present work provides an interferometric characterization of entangled neutral meson dynamics and highlights the role of entanglement in generating nonadditive phase structures.

{\it Acknowledgement :} I wish to thank Prof. Utpal Sarkar and Prof. Arghya Taraphder for support and encouragement. I would like to thank MHRD, Government of India for the research fellowship.

\end{document}